\documentclass[sigconf]{acmart}

\AtBeginDocument{%
  }

\copyrightyear{2025}
\acmYear{2025}
\setcopyright{rightsretained}
\acmConference[CHI EA '25]{Extended Abstracts of the CHI Conference on Human Factors in Computing Systems}{April 26-May 1, 2025}{Yokohama, Japan}
\acmBooktitle{Extended Abstracts of the CHI Conference on Human Factors in Computing Systems (CHI EA '25), April 26-May 1, 2025, Yokohama, Japan}\acmDOI{10.1145/3706599.3707399}
\acmISBN{979-8-4007-1395-8/25/04}

\begin{document}

\title[Optimizing SIA Development: A Case Study in User-Centered Design for Estuary, a Socially Interactive Agent Framework]{Optimizing SIA Development: A Case Study in User-Centered Design for Estuary, a Multimodal Socially Interactive Agent Framework}

\author{Spencer Lin}
\email{linspenc@usc.edu}
\orcid{0009-0007-0974-3395}
\affiliation{
  \institution{University of Southern California}
  \city{Los Angeles}
  \state{California}
  \country{USA}
}

\author{Miru Jun}
\authornote{Both authors contributed equally to this research.}
\email{mdjun@usc.edu}
\author{Basem Rizk}
\authornotemark[1]
\email{brizk@usc.edu}
\affiliation{
  \institution{University of Southern California}
  \city{Los Angeles}
  \state{California}
  \country{USA}
}

\author{Karen Shieh}
\email{karen93@berkeley.edu}
\affiliation{
  \institution{University of California, Berkeley}
  \city{Berkeley}
  \state{California}
  \country{USA}
}

\author{Scott Fisher}
\email{scott.fisher@usc.edu}
\affiliation{
  \institution{University of Southern California}
  \city{Los Angeles}
  \state{California}
  \country{USA}
}

\author{Sharon Mozgai}
\email{mozgai@ict.usc.edu}
\affiliation{
  \institution{University of Southern California}
  \city{Los Angeles}
  \state{California}
  \country{USA}
}


\begin{abstract}
    This case study presents our user-centered design model for Socially Intelligent Agent (SIA) development frameworks through our experience developing Estuary, an open source multimodal framework for building low-latency real-time socially interactive agents. We leverage the Rapid Assessment Process (RAP) to collect the thoughts of leading researchers in the field of SIAs regarding the current state of the art for SIA development as well as their evaluation of how well Estuary may potentially address current research gaps. We achieve this through a series of end-user interviews conducted by a fellow researcher in the community. We hope that the findings of our work will not only assist the continued development of Estuary but also guide the development of other future frameworks and technologies for SIAs.
\end{abstract}



\begin{teaserfigure}
  \includegraphics[width=\textwidth]{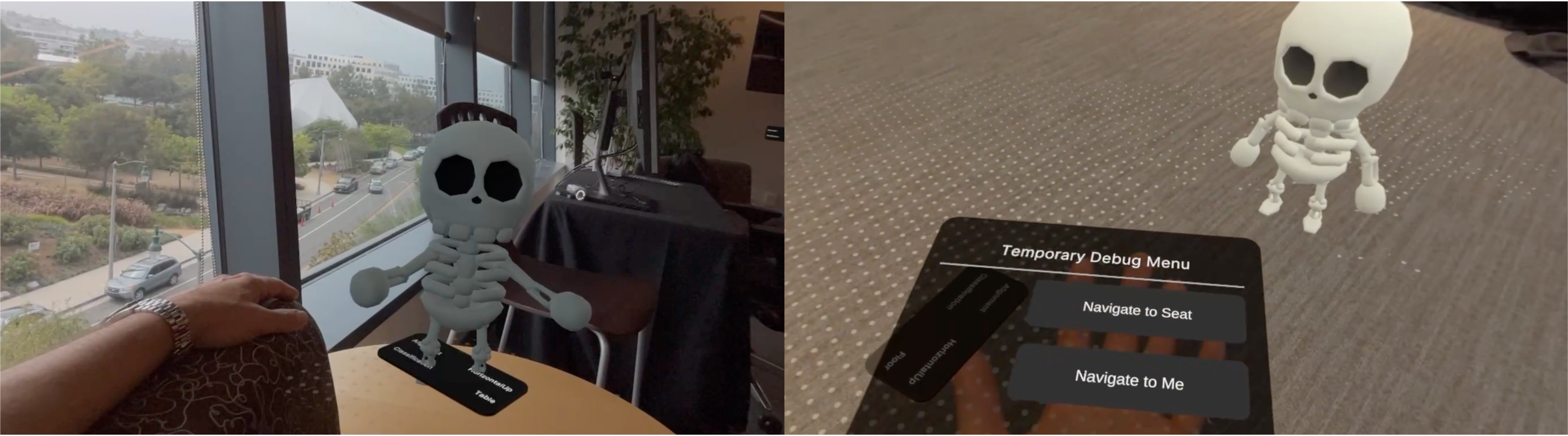}
  \caption{The Estuary demo featured an interactive cartoon agent in Augmented Reality on the Apple Vision Pro. [Left] Participant interacts with the agent after it jumps onto a detected seat. [Right] The hand-tracked menu triggers multimodal interactions.}
  \label{fig:teaser}
\end{teaserfigure}


\maketitle

\section{Introduction}

The increasing capability and widespread availability of artificial intelligence (AI) technologies have facilitated their use in the field of Socially Interactive Agents (SIAs), virtually embodied agents that are capable of autonomously communicating with people and each other in a socially intelligent manner using multimodal behaviors \cite{10.1145/3477322}. Despite significant interest in integrating modern AI-powered components into real-time SIA research, substantial friction persists due to the lack of a standardized, universal SIA framework. Developing an SIA requires integrating diverse technologies, including game engines as well as microservices to handle Voice Activity Detection (VAD), Automatic Speech Recognition (ASR), Large Language Models (LLMs), and Text-to-Speech (TTS), among others, all of which must work in harmony to enable seamless interactions \cite{10.1145/3477322}.  Due to the sheer breadth of technologies involved, it is inherently difficult to integrate and maintain them in a project. Oftentimes, this results in mundane repeat work between studies, performance compromises such as high latency due to the use of several cloud APIs, and/or incompatibilities of AI models due to hardware architecture mismatches or compute constraints\cite{hartholt2024multidisciplinary}\cite{rekik2024survey}. This is further exacerbated by an overall shift in interest from the field\cite{mozgai2023machine}\cite{pelachaud2021multimodal} towards multimodal processing, which not only requires low-latency response times but even further integration between modules and greater compute requirements.


To address these challenges, we developed Estuary \cite{estuary_iva2024}, an open-source multimodal framework for building low-latency real-time SIAs. Estuary's design is built on five key principles: 1) having the flexibility of interoperable modules, 2) being completely platform agnostic, 3) having the ability to be run off-cloud, 4) supporting multimodal processing, and 5) being open-source to the research community to facilitate the development and advancement of SIA technologies. 
The key objectives of this case study are to identify 1) What aspects of current state-of-the-art (SOTA) tools for developing SIAs do leading researchers in the field find most effective and valuable? 2) What critical limitations or gaps do SIA researchers identify in existing development tools that hinder the advancement of SIAs? 3) Which specific features, capabilities, or design principles of Estuary do SIA researchers perceive as most promising for addressing current challenges in the field? And finally, 4) What potential drawbacks or areas for improvement do SIA researchers identify in Estuary's approach to developing socially interactive agents? We used the Rapid Assessment Process (RAP) to conduct 1-hour informal interviews with SIA researchers, gathering insights on the state-of-the-art and feedback on Estuary \cite{BeebeRAP2005}. This analysis aims to inform the development of Estuary and future SIA frameworks.  

\section{Related Work}
We conducted a scoping review to systematically assess the current landscape of socially interactive agent (SIA) research from 2019 to 2024, with the specific aim of identifying studies that incorporate user-centered design principles in the development of SIA frameworks. We searched 8 electronic databases listed in Appendix ~\ref{searched_databases}, focusing on the past 5 years to align with the recruitment criteria of our study. The inclusion criteria for this review encompassed journal articles, conference proceedings, grey literature such as dissertations and theses, and review articles that were published in English. The search strategy was not limited by study design. Various search terms for SIAs were employed: SIA, virtual humans, or embodied conversational agent (ECA) are all popular terms in the academic literature, while digital human is often used in industry. More details about this search can be found in the appendix. Based on our search, we were able to find case studies that evaluated agent design principles, but were unable to identify user-centered design case studies for SIA frameworks which involve researchers in the development process. Despite the growing number of frameworks suitable for SIA development in recent years, there remains a conspicuous research gap in that there has yet to be an in-depth analysis of the actual needs and wants of SIA researchers. This research gap can be quite problematic, possibly leading to the development of frameworks that do not align with the needs of SIA research or which are simply unoptimized for SIA development. Hence, it is crucial to involve SIA researchers through case studies in user-centered design to better inform the direction of future SIA framework development efforts.

We then carried out a detailed analysis of three key frameworks and platforms—Virtual Human Toolkit (VHToolkit), Pipecat, and NVIDIA Avatar Cloud Engine (ACE)—to assess their respective strengths and limitations in supporting SIA development. These frameworks and platforms represent a range of realistic approaches, from academic research-oriented frameworks (VHToolkit and Pipecat) to commercial, enterprise-scale platforms (ACE), offering insights into the trade-offs researchers face when choosing between open-source frameworks and proprietary platforms. Given the diversity of use cases and technical requirements in the SIA field, no single framework or platform has emerged as a universally accepted standard, further justifying the need for comparative analysis. 

In addition to these, a final approach involves manually integrating cloud-based microservices, bypassing the use of a formal SIA development framework or platform. For instance, OpenAI's recently launched Realtime API enables developers to quickly integrate a fully-fledged voice-based conversational agent through a consolidated cloud endpoint, which may reduce latency. However, relying solely on cloud endpoints presents challenges such as reduced control over individual microservices and limitations in customizability. Moreover, managing multiple cloud microservices without a unified framework introduces complexities, including increased network propagation delays.

VHToolkit~\cite{hartholt2013all} is one of the most used and enduring research SIA frameworks across domains including affective research in healthcare~\cite{devault2014simsensei}, cognitive science research~\cite{gratch2013virtual}, and intelligent tutoring~\cite{nye2017building, swartout2016designing}. It helps facilitate SIA development by providing a scalable modular Unity project. Even so, several gaps exist in this system such as lacking off-cloud capabilities when leveraging modern AI technologies. This makes it difficult to conduct studies when internet access is limited, privacy of conversations is important, or when reproducibility and sustainability are important factors seeing as cloud services frequently change. Finally, the nature of VHToolkit being coupled with Unity makes it difficult to integrate open-source AI models from the research community which are commonly written in Python. 

Pipecat \cite{Flaqué2024pipecat} is a recent open-source framework for building voice and multimodal conversational AI. It provides convenient modular wrappers for various microservices ranging from ASR, TTS, and LLMs to quickly set up a conversational bot for a business use-case such as customer support. The advantage of using Pipecat is its universality in integration with popular microservices. Moreover, it can run as a backend and connected to an SIA application as a singular, consolidated endpoint which shifts the work of optimizing and maintaining microservices away from researchers. One major drawback is the lack of integration with common game engines for developing SIAs such as Unity, which greatly nullifies its theoretical advantages as a consolidated endpoint. Furthermore, Pipecat harnesses the same cloud microservices (e.g., ElevenLabs, OpenAI), which does not provide any additional privacy, latency improvements, cost savings, or flexibility compared to more established research frameworks such as VHToolkit.


NVIDIA Avatar Cloud Engine (ACE) \cite{nvidia_ace} is a commercial platform which provides a suite of pre-built customizable microservices ranging from NVIDIA proprietary technologies to NVIDIA-optimized versions of open-source LLMs.  
They are part of the more general NVIDIA Unified Compute Framework (UCF) \cite{nvidia_ucf}, which aims to deliver easy-to-deploy AI products that require little coding experience. Due to its ability to be run as an off-cloud solution, it can have much lower latency by incurring zero network propagation delay. NVIDIA ACE also features a graphical user interface (GUI) for building SIAs which improves accessibility. However, NVIDIA ACE is geared towards enterprise-scale applications with cost-prohibitive licensing, the need for private consulting if an on-prem (off-cloud) solution were to be set up, and finally perhaps most limiting, NVIDIA ACE only works comprehensively with their proprietary Omniverse engine. In short, it is difficult for researchers to leverage the power of this framework as it is not open, has a high barrier of entry, and relies on uncommon tools such as Omniverse which the SIA field has yet to adopt.

Given the current landscape of frameworks, there is a need for a more modern framework that can adequately address the research gaps that have formed since the advent of AI-powered SIA microservices. Integration with existing SIA research and tools, interoperability of microservices, off-cloud, open-source, and accessibility are all evidently important qualities in an SIA framework. Hence, these gaps motivate the creation of a new framework (Estuary) not only architected with those qualities in mind but built with the input of researchers throughout each step of the way.

\section{Estuary: A Multimodal SIA Framework}
Estuary is an open-source multimodal framework for building low-latency
real-time SIAs. It provides several advantages such as being able to be run entirely off-cloud; allowing data to remain private, SIAs to continue to operate in offline environments, and for studies to be more reproducible as cloud services often change. Employing a client-server-based architecture allows computations to be run on-edge. This enables researchers to overcome hardware platform issues such as hardware architecture incompatibilities or compute constraints. This is especially useful when deploying off-cloud SIAs on non-NVIDIA (many AI models rely upon NVIDIA hardware to function) and computationally limited devices such as Extended Reality (XR) headsets. This opens the door for more multimodal capabilities that leverage XR hardware. Currently, Estuary is integrated with XR libraries such as ARKit\cite{UnityARKitDocs} which gives SIAs the ability to semantically understand the user's vicinity and interact with the user in AR. Finally, as an open-source framework with an interoperable microservice design, users can freely use or integrate their own microservices while performing data analysis at any point in the pipeline without the restrictions often imposed by proprietary platforms.

\begin{figure}[h]
    \centering
    \includegraphics[width=1\linewidth]{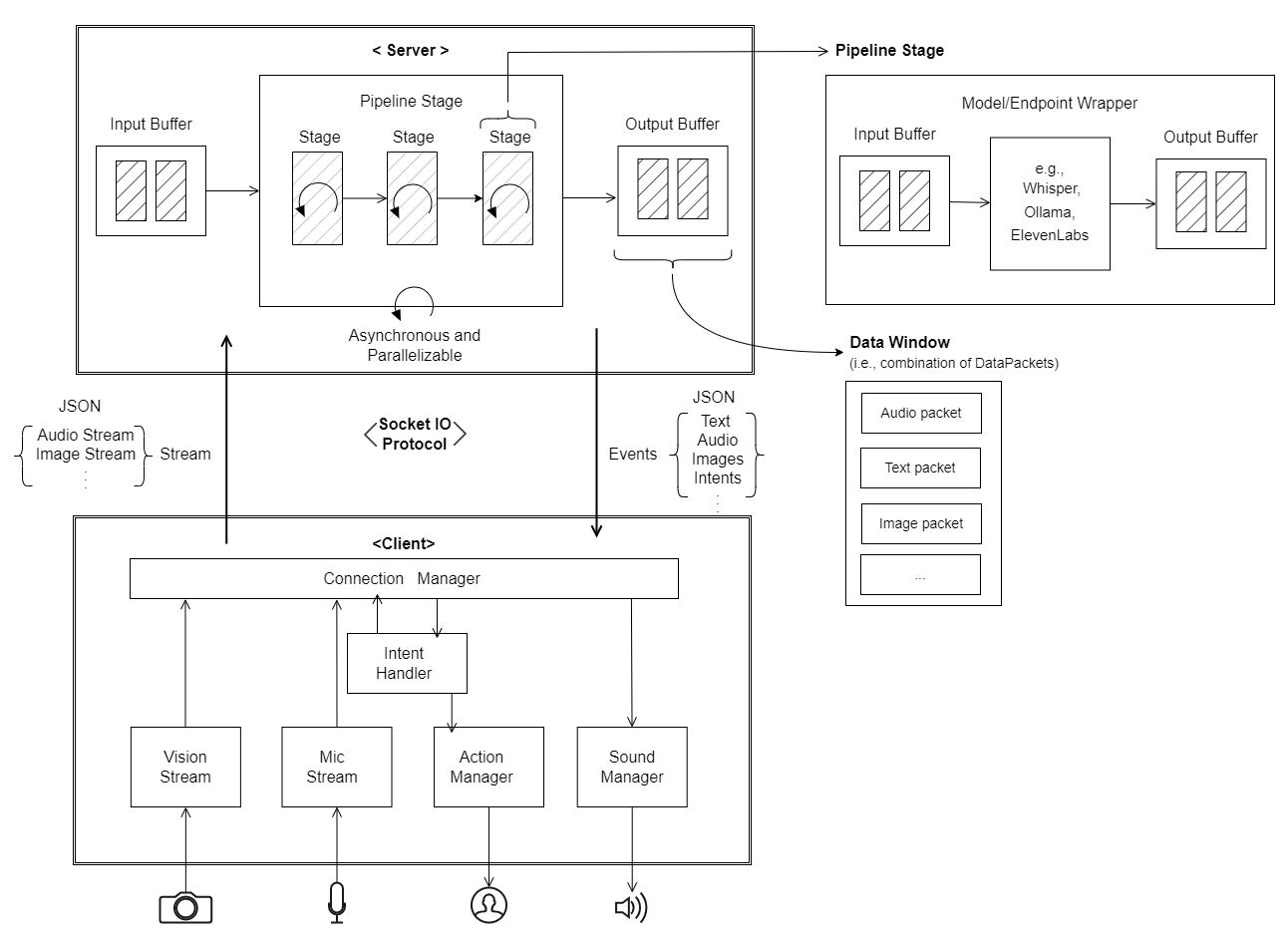}
    \caption{Abstract Diagram of Estuary System Architecture: A client-server based architecture employing $SocketIO$ Protocol, where the data is transferred in $JSON$ format, while it gets parsed and defined according to defined event types. The two boxes on the right illustrate the versatility of Estuary to extend to any type of modalities according to the DataPackets design, as well as to support any model or online service endpoint by extending/configuring with minimal code the appropriate abstract Stage definition.}
    \label{fig:estuary_system_arch}
\end{figure}

The framework core, illustrated in the <Server> box in Figure \ref{fig:estuary_system_arch}, has a highly modular design, where the different components (e.g., VAD, TTS) would be running independently wrapped in what is referred to as "Stages." These Stages are designed for asynchronous, parallel execution, with each Stage handling the inference call in an isolated child process. Inputs are unpacked according to the defined types of the stage input and outputs, aggregated, then processed, and the outputs are dispatched in a similar fashion. Through a connected user-defined sequence of Stages, referred to as a Pipeline, the system orchestrates data flow using any amount of data, containing multiple diverse DataPackets which can support multimodal information (e.g., $AudioPacket$, $TextPacket$, $ImagePacket$ etc.). This implementation allows Estuary to support a wide variety of multimodal modules in the future.

The client-server architecture as shown in Figure \ref{fig:estuary_system_arch}, employs a simple event-based $SocketIO$ protocol \cite{socket_io}, where the server runs in Python while the client can be implemented in any other languages with $SocketIO$ support. Moreover, the pipeline stage is designed according to an events-based life-cycle game loop fashion\cite{gregory2018game}. This makes it more analogous to and hence easier to interface with game engines. To further aid in development, we created a Software Development Kit (SDK) which is integrated into Unity's native package manager which makes it easy to set up a client as well as example projects with pre-made clients for Apple Vision Pro (AVP), Meta Quest 3, and a standalone desktop application. 

All in all, Estuary's research-focused design is intended to support the SIA research community by being open-source and future-facing with valuable features such as off-cloud capabilities, an interoperable microservice design, hardware platform agnosticism thanks to its client-server architecture, and support for multimodal processing.

\section{Methodology}

\subsection{Demographic and Recruitment}
The goal of the case study was to investigate the current need areas for SIA research tools in order to steer the development of Estuary. Therefore, we sought input from active members of the SIA research community representing diverse perspectives to ensure a comprehensive understanding of their needs. We identified a target demographic of researchers over 18 years old with direct experience developing virtual agents. To ensure relevancy, researchers must have participated in SIA research within the last five years. Furthermore, participants must have previously fulfilled the role of researcher, developer, or project manager involved in the conceptualization, design, or technical development of SIAs. 

To facilitate face-to-face interviews, we utilized a convenience sampling approach by recruiting participants who have been associated with the USC Institute for Creative Technologies (ICT) within the past 5 years. This approach was selected due to the accessibility of this population and their direct experience with the technologies under investigation. 
For recruitment, an internal email including a comprehensive informational sheet of the study was sent to prospective participants' ICT email addresses. The study was conducted on a first-come-first-serve basis. In accordance with the Rapid Assessment Process (RAP) protocol \cite{BeebeRAP2005}, our objective was to recruit 10 participants, as prior studies indicate that recurring themes typically emerge after conducting 8-10 interviews. 10 participants (n=3 women; n=7 men) volunteered and completed our study. Of these participants, there were researchers (n=5), a developer (n=1), research \& development (r\&d) staff (n=2), and PhD students (n=2). Among the researchers and research \& development staff, n=4 and n=1 hold a doctoral degree respectively.  

\subsection{RAP Protocol}
We chose the RAP protocol for this case study to ensure we could gather rich, insider perspectives from participants, tailored to meet the unique demands of SIA research. The RAP method fosters "directed conversations" that encourage thoughtful and context-specific responses while allowing the interviewer, who must be an insider in the community, to guide the conversation with precise, domain-relevant vocabulary. Furthermore, the RAP protocol's emphasis on rapid yet comprehensive data collection and analysis, conducted by at least two researchers with multidisciplinary expertise, allowed us to efficiently capture the nuanced insights we needed. This was particularly important for our study, as it ensured that the interviewer, also an active SIA researcher at ICT, could elicit highly relevant data and engage deeply with the subject matter.

\subsection{Data Collection}
The interviews were one-on-one and approximately one hour in length. All interviews were recorded via video and audio for future data analysis by team members. Participants were welcome to provide comments or feedback at any time during the session and both the interviewer or the participant were allowed to engage with follow-up questions. The interviewer made sure to solicit a response from each participant for each of the questions outlined in the interview questions in Appendices~\ref{baseline_questions} and ~\ref{estuary_questions}.
During each session, participants answered baseline questions to establish their background and views on the SOTA for SIA development, were briefed on what Estuary is, participated in a demonstration of Estuary on the AVP headset, and finally answered questions about their experience with Estuary. A more detailed description of the study procedures can be found in Appendix~\ref{interview_guide}.

The Estuary demonstration featured an Augmented Reality (AR) application running on the AVP headset, showcasing a small, interactive cartoon character, as seen in Figure~\ref{fig:teaser}. The character was able to navigate around real-world obstacles, jump on and off detected chairs, and hold a conversation, powered by ChatGPT 3.5\cite{gpt3.5}. We had originally planned for the character to have Natural Language Understanding (NLU) capabilities so that it may understand commands such as "come sit with me" or "come to me" and pathfind to the target destination accordingly. As NLU capabilities were not yet implemented, we built an AR menu as seen in Figure ~\ref{fig:teaser} that tracks the user's hand so users can trigger the character's multimodal interactions. The menu consisted of two buttons for the character to navigate to the user and navigate and jump onto the closest seat to the user. Participants were shown how to use the hand menu and instructed to converse with the character. Once familiarized, they were given time to freely interact with the character. Though this is a simple early-stage application, it serves as a technical demonstration of all the features Estuary currently has, including real-time voice interactions, spatial semantic understanding of the AR environment, and dynamic pathfinding in AR. This provided a representative proof-of-concept to ground discussions in our study. Furthermore, this allowed us to involve researchers early in the development of Estuary to iterate using input from the research community.

\subsection{Thematic Analysis}
Four research questions (RQs) were identified: 1) What aspects of current state-of-the-art tools for developing Socially Interactive Agents (SIAs) do leading researchers in the field find most effective and valuable? 2) What critical limitations or gaps do SIA researchers identify in existing development tools that hinder the advancement of socially interactive agents? 3) Which specific features, capabilities, or design principles of Estuary do SIA researchers perceive as most promising for addressing current challenges in the field? 4) What potential drawbacks or areas for improvement do SIA researchers identify in Estuary's approach to developing socially interactive agents? Guided by the research questions, we generated transcripts of the interviews using OpenAI's Whisper Turbo \cite{radford2023robust}, coded the transcripts using NVivo 15 software, and conducted thematic analysis on the subsequent groups of codes. Two members of the research team individually coded 5 transcriptions each, which were then cross-reviewed and edited as necessary. The codes were emergent from the transcriptions rather than predetermined and were categorized under the research questions. After all transcripts were coded, the codes were grouped into broader themes, which were then analyzed and visualized \cite{miles1994qualitative}.


\section{Findings}

Before and after introducing Estuary, participants were asked guiding questions to identify the values and gaps in SIA development tools and Estuary. A sample of interview excerpts can be found in Appendix~\ref{interview_excerpts}. From their responses, several key themes emerged for each of the research questions.
\subsection{RQ1 - Current Valuable Aspects of SOTA Tools for Developing SIAs}

\textbf{Advanced Language Capabilities and Performance. } Most participants (n=7) highlighted the significant progress that LLMs have brought to SIAs; enhancing language comprehension, enabling more personalized interactions, and simplifying the setup of agents through prompt engineering. To underscore the importance of LLM quality, some (n=3 researchers) even mentioned they would prioritize developing a more performant agent over graphical fidelity.

\textbf{High-Quality Multimodal Interaction and Representation. } Half of the participants (n=5) remarked on how modern tools have improved multimodal interaction and representation of SIAs. This group of participants includes r\&d staff (n=2), PhD students (n=2), and a researcher (n=1), representing a broad sentiment across multiple roles. They outlined qualities such as high-fidelity graphics and lip-sync, responsive and accurate ASR and TTS, and multimodal emotional perception to all contribute towards more life-like SIAs. Altogether, many of these key features are now quite robust. They are able to provide versatile functionality for a variety of research studies and have become more accessible to diverse users -- perfect English is not required for ASR to work well.

\textbf{Ease of Integration and Interoperability. } Another salient theme that emerged was the growing ease with which modern platforms facilitate the integration of key microservices for building SIAs such as ASR, LLMs, TTS, and lip-sync. One researcher (n=1) and one r\&d staff member (n=1) highlighted companies such as Unity, a popular game engine that supports many third-party microservice integrations, and NVIDIA, which has been developing accessible microservices that are integratable through intuitive APIs. In addition, one developer (n=1) mentioned that the interoperability of microservices also enhances the robustness of SIAs, as microservices can easily be swapped out when issues arise.

\textbf{Usability and Accessibility of Development Tools. } A distinct theme that emerged was the general increasing usability and accessibility of development tools for SIAs. Several researchers (n=2) and one r\&d staff member (n=1) emphasized how tools like character creation applications, prompt-driven LLMs, and cloud-based services have simplified the development process, especially for researchers coming from non-computer science backgrounds. These advancements not only make SIA creation more intuitive, but also reduce the burden of local computing resources, allowing researchers and developers to focus more on design and functionality.

\subsection{RQ2 - Current Gaps in SOTA Tools for Developing SIAs}
\textbf{Challenges with Integration. } Most participants (n=8) highlighted the difficulty of using SOTA tools, with technical integration issues being a major challenge for participants across all roles (n=2 researchers, n=2 r\&d staff, n=1 developer, n=1 PhD student). While there have been significant advancements to each individual microservice, they are often isolated on different platforms with incompatible architectures and varying software and hardware requirements, requiring teams to be assembled with specialized technical skills to bridge these gaps. For example, a proprietary lip-sync microservice offered by NVIDIA ACE may not be compatible with Unity. Furthermore, many off-the-shelf microservices may not be fine-tunable to the degree researchers need and/or have inadequate API documentation, further complicating integration efforts. On the other hand, the majority of the same participants as well as two other researchers (n=3 researchers, n=1 developer, n=1 r\&d staff, n=1 PhD student) expressed a strong desire for tools that enhance user experience by automating the complex, technical steps involved in integrating microservices.

\textbf{Technical and Performance Issues. } All but one participant (n=9) identified technical and performance issues with current tools. The most widespread concern (n=8) focused on the intelligence and reliability of LLMs. Participants expressed a strong need for controllable LLM behavior to ensure accurate information, preservation of meaning, and emotional sensitivity, particularly for vulnerable populations. Although fine-tuning models can improve reliability for specific use cases, this is often impractical due to the limited data available to individual researchers and the difficulty in developing adequate guardrails. Additionally, concerns arose about latency, noting that delays between a user’s utterance and the agent's response can hinder real-time interaction, which is critical for a responsive and engaging SIA.

\textbf{Multimodal Interaction Gaps. } Four participants across most of the roles (n=2 researchers, n=1 r\&d staff, n=1 PhD student) pointed out a gap in multimodal interactions, with three specifically emphasizing the importance of nonverbal behavior in SIAs. While they acknowledged improvements in technologies like lip-sync, they noted that these systems still feel unnatural, and generated gestures often fail to align with the content of speech. Of this group, three participants identified other missing multimodal features, particularly the integration of audiovisual data. They highlighted the need for more versatile tools that can seamlessly analyze inputs such as emotions and fatigue to enhance overall interactions. 

\textbf{Privacy. } Two participants (n=1 developer, n=1 researcher) raised important concerns about privacy and security issues related to cloud-based microservices. They emphasized that conversations with LLMs may involve sensitive personal information, which individuals may not want stored or accessed by third-party companies. In addition, they highlighted significant security risks, particularly in projects involving the US Army, where safeguarding information is critical. This theme underscores the need for more secure, privacy-conscious solutions to protect both personal data and classified information.

\textbf{Sustainability. } Sustainability concerns were highlighted by r\&d staff (n=2), a developer (n=1), and a researcher (n=1), focusing on challenges such as limited funding, the high cost of maintaining individually developed research tools, and the rapid evolution of standards and technologies. Participants noted that keeping tools updated requires significant resources, which are rarely supported by research funding. Similarly, with commercial tools, constantly changing standards and API versions make it difficult for researchers to maintain stable pipelines throughout the duration of their studies. These ongoing challenges create additional overhead for development and ultimately impede progress in the field.

\subsection{RQ3 - Valuable Aspects of Estuary}

\textbf{Flexibility and Versatility. } Participants responded positively to Estuary's flexible design philosophy. Most participants (n=8), with the exception of a developer and a researcher, appreciated the option to run SIAs either entirely off-cloud or by leveraging cloud services, depending on the use case. The off-cloud capability was seen as particularly valuable for applications in therapeutic or military contexts, where privacy is critical, or in environments with limited or no internet connectivity. Conversely, participants also recognized the advantages of cloud-based microservices for simplifying the development process. Additionally, most participants (n=7) emphasized the importance of Estuary's client platform-agnostic nature, particularly its support for XR devices, which enables SIAs to operate in shared physical spaces with human users.

\textbf{Scalability and Future-Facing Architecture.} Four participants (n=2 researchers, n=2 r\&d staff) gave positive remarks on Estuary's architectural design. Due to Estuary taking an edge computing approach, it allows more capable SIAs to run on smaller devices while being designed for more multimodal capabilities to be integrated in the future such as computer-vision-driven AI models. This is aided by Estuary being written in Python and being architecturally designed around integrating with popular open-source libraries such as Hugging Face and Ollama.

\textbf{Efficiency and Usability in Low-Latency Design. } Two participants (n=1 researcher, n=1 PhD student) highlighted the value of Estuary’s efficiency and usability, particularly in terms of its low-latency verbal interaction. This was seen as a key benefit, likely due to Estuary's streamlined architecture, which reduces the number of cloud endpoints and minimizes propagation delays. Its design also enables efficient audio and data streaming between modules (e.g., LLM to TTS to client), allowing for simultaneous processing. Additionally, one developer (n=1) and one r\&d staff member (n=1) appreciated the ease of setup, noting the utility of the provided client SDK, which integrates smoothly with Unity’s native package manager. The use of a unified endpoint was also seen as a practical advantage, simplifying system management compared to platforms that require managing multiple endpoints for different microservices.

\textbf{Openness and Collaboration. } Estuary’s open-source nature was valued by (n=2) researchers and (n=1) r\&d staff member  who appreciated the flexibility it offers. They saw that this quality will allow them to modify and extend Estuary by incorporating other open-source AI microservices or by adapting it to work with existing SIA research frameworks like the Virtual Human Toolkit (VHToolkit). This adaptability was seen as a significant advantage, as it enables researchers to customize Estuary to suit specific project requirements, enhance functionality, and ensure better integration with their established workflows.

\subsection{RQ4 - Gaps in Estuary}
\textbf{Multimodal and Interaction Improvements. }
Many participants (n=7) suggested additional modalities to improve engagement and immersion within Estuary. Audiovisual enhancements were a key interest, with some proposing the integration of haptics and biometric signals to track gestures, facial expressions, eye movements, and analyze speech emotions and semantics. These capabilities could enrich AR interactions by triggering actions based on proximity, emotional states, or physiological data. Participants also highlighted the need for more natural nonverbal communication, such as turn-taking, back-channeling, and hand gesture controls, with a few advocating for longitudinal interaction support in which SIAs "remember" users across sessions. Additionally, there was interest in exploring multi-agent support through Estuary’s architecture, pointing to a desire for further expansion and integration.

\textbf{Integration and Expansion. }
Several participants (n=2 r\&d staff, n=1 developer) were keen on Estuary integrating more features as well as additional architectural and software design considerations for easy integration of past and future pipelines and technologies. The developer expressed interest in the integration of a dialogue manager for the option of foregoing an LLM entirely in the interest of additional controllability and reducing the complexity of the system. The two r\&d staff members expressed that they would ideally like greater flexibility in choice of client platforms such as web apps. Though Estuary is capable of supporting web clients, it would need to be expanded further so that it can be a practical and scalable solution that can be deployed to the cloud to support web apps.  


\textbf{Performance and Usability Concerns. }
A large number of participants (n=3 researchers, n=1 developer, n=1 r\&d staff, n=1 PhD student) had reservations about the usability of Estuary's client-server architecture, citing a complex setup process involving connecting multiple devices through a stable local network. Having as simple of a setup is understandably much preferred to improve usability, reliability, and reduce development overhead. Furthermore, one r\&d staff member had concerns about potential performance bottlenecks in the architecture which would introduce additional latency due to data congestion.



\section{Discussion}
This case study gathered feedback from a broad cross-section of members in the field of SIAs ranging from students to veteran researchers. By adhering to RAP, we aimed to collect high-quality data that reflects the real-world experiences and needs of the SIA research community, making it a suitable choice for guiding the future development of Estuary and related frameworks.
The thematic analysis done for this case study identified the main problems and needs for SIA frameworks. We thus propose the following key considerations for developing a comprehensive, flexible, and sustainable SIA framework:

\textbf{Open-Source. }
Research is a collaborative endeavor and so should the frameworks that support it. An open-source framework allows for more of the community to collectively contribute as well as nurtures adoption across the field. Furthermore, access to the source code allows researchers to build and integrate SIA microservices anywhere within the framework. It also guarantees longevity, ensuring that the framework will not abruptly cease to exist as it is not software as a service.

\textbf{Standardization \& integration with existing SIA frameworks. }
An SIA framework should ideally be sustainable such as being able to integrate with and extend upon existing SIA frameworks and their functionalities. One integration method that was suggested is standardizing the message protocol amongst microservices with existing frameworks. However, it should be recognized that enforcing a certain standard may limit the flexibility to integrate with other frameworks and/or limit future development to follow practices that may no longer be optimal. As such, the decision to integrate with existing standards requires a careful consideration of the trade-offs.

\textbf{Interoperable microservice architecture. } 
In conjunction with being open-source, an interoperable microservice architecture well-integrated with popular cloud APIs as well as model libraries such as Hugging Face allows for the easy integration of existing and future microservices. 

\textbf{Off-cloud capabilities. }
An SIA framework with off-cloud capabilities is crucial for scenarios that necessitate privacy, data security, offline operations, and/or reproducible results.

\textbf{Accessible GUI. }
Providing an accessible GUI was a consensus amongst the participants. This provides the benefits of boosting ease of use and supporting the broader diverse SIA community beyond just developers. 

\textbf{Complexity of a client-server architecture. } 
A prevalent concern amongst participants was that a client-server setup might be too complicated. As such, it is important that SIA frameworks with a client-server architecture can also be run standalone on one device.

\subsection{Limitations}
While the convenience sample allowed for efficient data collection from relevant experts, it may limit the generalizability of our findings. This case study focused on researchers at ICT, and future research should include a more diverse and geographically dispersed sample to enhance external validity. Participants often provided feedback on the demo rather than the framework itself, but this distinction became clearer after adjusting the interview protocol to request specific feedback on both. Since Estuary is still in early development, with experimental off-cloud LLM support, future studies should ideally use fully off-cloud microservices to better represent the framework.


\subsection{Future Research}
We plan to incorporate several key features into Estuary's development road map. This includes expanding sensor integration, such as front-camera access on XR devices, and adding NLU capabilities to enhance multimodal interactions. Additionally, we aim to develop a user-friendly GUI to make Estuary accessible to a wider range of researchers, particularly those without coding expertise. To support the open-source community, we will provide comprehensive documentation and tutorials, fostering a collaborative environment. Lastly, improvements to the client-server architecture are expected to significantly reduce latency. The interviews conducted through the RAP process were immensely helpful in shaping Estuary's road map. Thus, after implementing the requested features, we plan on collecting further feedback through iterative rounds of RAP. Considering how conducive the demo was for eliciting salient feedback, we hope to better showcase the new features in either a unified demo or a series of demos focusing on specific features.

\begin{acks}
Part of the efforts depicted were sponsored by the US
Army under contract W911NF-14-D-0005. The content of
the information does not necessarily reflect the position or
the policy of the Government, and no official endorsement
should be inferred. 
\end{acks}

\bibliographystyle{ACM-Reference-Format}
\bibliography{references}

\appendix

\section{Database Information}
\subsection{Searched Databases}
\label{searched_databases}
\begin{itemize}
    \item ACM Digital Library
    \item IEEE Xplore
    \item PubMed
    \item Google Scholar
    \item SpringerLink
    \item Scopus
    \item PsycINFO
    \item Web of Science
\end{itemize}

\subsection{Database Search Terms}
\label{database_search_terms}
\begin{itemize}
    \item “Socially Interactive Agents” AND “end-user evaluation”
    \item “Virtual Humans” AND “end-user experience” OR “user interaction”
    \item "Socially intelligent agents” AND “usability testing”
    \item “Conversational agents” AND “user satisfaction” OR “user engagement”
    \item “Embodied conversational agents” AND “user-centered design”
    \item “SIA frameworks” AND “usability study” OR “user acceptance”
    \item “Multimodal agents” AND “user perceptions”
    \item “Digital Humans” AND “usability-study”
\end{itemize}

\section{Estuary Architecture}

\section{Interview Overview}
\subsection{Interview Guide}
\label{interview_guide}
The general overview that the interview adhered to.
\begin{enumerate}
    \item Press record on camera recorder
    \item Read the script and introduce the study
    \item Ask the baseline interview questions
    \item Give a brief, high-level overview of Estuary
    \item Start a Guest Session and mirror the display of the AVP to a MacBook for monitoring purposes
    \item Start a screen recording on the machine running the Estuary server
    \item Start a screen recording on the MacBook
    \item Guide the participant through the AVP calibration process
    \item Have the participant demo the AVP Estuary client
    \item Ask the Estuary focused interview questions
    \item Conclude the study
\end{enumerate}

\subsection{Baseline Questions}
\label{baseline_questions}
These were the questions asked by the interviewer which provided the baseline for SIA current tools and trends.
\begin{enumerate}
    \item Could you briefly describe your experience in SIA research?
    \item Please describe your experience with XR devices.
    \item What are some tools you use for developing SIAs? (e.g. Character Creator? ASR?  LLMs?  TTS?)
    \item Could you please describe what you view as the current state of the art for SIA development and what you think about it?
    \item What are some things that you feel the state of the art is lacking?
    \item What do you like and dislike about the SIA development process?
    \item What hardware platforms do you currently use for SIAs? Are there any hardware platforms you would like to use but cannot for whatever reason and what are they?
    \item In a similar vein, have you ever run into any moments where you wish you could use a specific piece of software or AI model (e.g. ASR, TTS, LLM, etc.) but you couldn’t?  If so, what was that software and why couldn’t you?
    \item What would you say are some things on your wishlist in terms of SIA research?
\end{enumerate}

\subsection{Estuary Questions}
\label{estuary_questions}
These were the questions asked by the interviewer which provided feedback from the participants for Estuary.
\begin{enumerate}
    \item What was your overall impression of Estuary?
    \item Did you have particular likes or dislikes?
    \item What are some of Estuary’s most appealing features to you?
    \item What are some of Estuary’s key limitations you see?
    \item What are your thoughts on our approach of offsetting heavy computations to a separate networked computer rather than on one standalone device?  Would you prefer running everything on one device or delegating heavy computations to a separate device?
    \item Do you foresee yourself ever running an SIA research study that uses purely off-cloud processing ie. ASR, LLM, TTS?  If not, what percentage of off-cloud / cloud APIs would you leverage and why?
    \item Would you in the future consider leveraging the AR capabilities of Estuary?  Why or why not?
    \item Would you use open-source AI models ie. from HuggingFace in an SIA or be interested in running your own AI model on an SIA?  What sort of model and how adequately do you feel Estuary can fulfill that need? 
    \item Is multimodal processing of interest to you and your research? What modalities of input are you interested in, and what kind of multimodal processing would you like to use?
    \item Would you use Estuary in a future research study, if yes, how would you?
    \item Compare and contrast Estuary with other similar tools: what are some value-adds and what can be improved?
    \item What are some features and capabilities you would like to see incorporated in the future?
    \item What are some other suggestions you have for improvement if any?
    \item Is there an idiosyncratic, such a far-out sci-fi-esque vision of yours with SIAs that you would just love to execute upon and what is that?
    \item Is there anything else you would like to add?

\end{enumerate}

\section{Interview Excerpts}
\label{interview_excerpts}
\begin{table}[ht]
    \centering
    \begin{tabular}{p{0.2\linewidth} | p{0.75\linewidth}}
        Themes  & Interview Excerpts \\ \hline \hline
        \vspace{0.5pt}
        RQ1 - Advanced AI Capabilities and Performance & 
            \vspace{0.5pt}
            \textit{
            With ChatGPT models, you can sort of push the boundaries further and create characters that have a very wide sort of – they can talk about many different things, like a very wide field domain. It's like open world kind of interaction.} \\

        \vspace{0.5pt}
        RQ2 - Sustainability & 
            \vspace{1.0pt}
            \textit{People still want to come do
            demos of SimSensei, and it doesn't exist anymore because, things like that, research was done
            in 2017 and 2018, and literally technology updates so fast.} \\

        \vspace{0.5pt}
        RQ3 - Platform Flexibility & 
            \vspace{1.5pt}
            \textit{
            I think that the most compelling reason to run your own models,
            or at least offline models, is what you mentioned, where there's consistency. So for that, you would want
            to have more control over the models, and then the trade-off clearly is that oftentimes
            the more powerful models are cloud-based, so I think having that, that choice is, is helpful.} \\

        \vspace{0.5pt}
        RQ4 - Multimodal and Interaction Improvements & 
            \vspace{0.5pt}
            \textit{I want it all. I want it all. I want to be able to track facial expression of a user. I want to be able to capture the semantic nature of their speech, the acoustic properties of their speech, their behaviors, their movements, gestures, their biosense signals.
            You know, I want the ultimate Skinner box. You know, I want controlled stimulus environment where I can measure a whole spectrum of elements, eye tracking, pupillary dilation, heart rate, skin conductance.} \\

    \end{tabular}
    \caption{Samples of interview excerpts for selected codes.}
    \label{tab:excerpts}
\end{table}

\end{document}